\documentclass[11pt]{article}
\usepackage[a4paper,margin=2.5cm]{geometry}
\usepackage{amsmath,amssymb}
\usepackage{graphicx}
\usepackage{adjustbox}
\usepackage{float}
\usepackage{booktabs}
\usepackage{microtype}
\usepackage{siunitx}
\usepackage[hidelinks]{hyperref}
\title{Atmospheric Newtonian noise in torsion-balance measurements of the gravitational constant $G$}
\author{Jyotirmaya Mohanta$^{a}$ and Yutaka Shikano$^{b,c,d}$\footnote{Corresponding author: \texttt{yshikano@cs.tsukuba.ac.jp}.} \\
$^{a}$Graduate Program in Computer Science, University of Tsukuba, Japan \\
$^{b}$Institute of Systems and Information Engineering, \\ University of Tsukuba, Tsukuba, Ibaraki 305-8573, Japan \\
$^{c}$Center for Artificial Intelligence Research, \\ University of Tsukuba, Tsukuba, Ibaraki 305-8577, Japan \\
$^{d}$Institute for Quantum Studies, Chapman University, Orange, CA 92866, USA}
\date{}

\begin{document}
\maketitle

\begin{abstract}
Measurements of Newton's gravitational constant remain limited by environmental and apparatus-dependent systematics whose treatment is often less explicit than that of instrumental noise. Among these, atmospheric density fluctuations generate unshieldable gravity gradients that couple directly to torsion-balance observables as atmospheric Newtonian noise. Here, we develop a GUM-consistent framework for propagating this contribution through the torque estimator into the uncertainty budget of torsion-balance measurements. We derive closed-form spatial transfer functions for two benchmark geometries: a two-mass dumbbell, which provides an upper-coupling reference, and a perfectly symmetric cross, which serves as an idealized rejection limit and exposes the trade-off between suppressing low-order environmental coupling and preserving signal response. We also separate stationary correlated inputs from non-stationary baseline drift, using the Ornstein--Uhlenbeck process only as a benchmark for the former. Atmospheric pressure benchmarks indicate that the resulting background contribution is below present reference uncertainty levels, but can become relevant as systematic floors approach the part-per-million regime. This framework provides a practical route for incorporating site-specific environmental gravity gradients into future torsion-balance uncertainty budgets.
\end{abstract}


\section{Introduction}
The measurement of Newton's gravitational constant $G$ remains challenging: the recommended value still carries a relative standard uncertainty of order $10^{-5}$~\cite{CODATA2022,NISTCODATA}.
In this regime, metrologically complete uncertainty budgets must address not only instrumental noise but also subtle environmental couplings.
Environmental influences (seismic, barometric, hydrological, and building-related) are discussed explicitly in a number of determinations and reviews~\cite{LutherTowler1982,Schlamminger2002,Speake2005}.

While NN is unlikely to explain the historical scatter in published $G$ values by itself, its metrological relevance increases as experiments push toward relative uncertainties at or below $10^{-6}$.
Instrumental (thermal) noise typically averages down as $T^{-1/2}$ for a simple mean estimator, whereas environmental Newtonian fields can exhibit long correlation times and therefore contribute persistent variance or bias to long integrations.
The atmospheric benchmark developed below (Sec.~\ref{sec:atmos} and Fig.~\ref{fig:pressure_to_torque}) implies sample-mean equivalent gradient uncertainties that are typically around $10^{-16}$--$10^{-14}$~\si{\per\square\second} for $T\sim10^{3}$--$10^{5}$~s at $z_0\sim\SI{1}{m}$, depending on the chosen pressure envelope and the benchmark $k$--$\omega$ mapping.
For the two illustrative benchmark signal gradients introduced later in Sec.~\ref{sec:benchmark_gamma}, $\Gamma_{\rm sig}=10^{-7}$ and $10^{-6}$~\si{\per\square\second}, this corresponds to relative levels below the current $\sim 2.2\times10^{-5}$ CODATA relative standard uncertainty, though potentially relevant for next-generation targets below $10^{-6}$ and for controlling Type-B (systematic) effects.
The present paper therefore does not claim that atmospheric NN dominates any particular historical experiment; rather, it provides a benchmark framework showing how to judge relevance once an apparatus-specific field model, transfer function, and estimator are supplied.

Environmental density perturbations produce fluctuating gravitational forces and gradients that cannot be shielded.
In the gravitational-wave community these effects are known as Newtonian noise and have been studied extensively for interferometric detectors~\cite{Creighton2008,Coughlin2018}.
For torsion-balance measurements of $G$, the physics of NN coupling is closely related, but its role in the measurement problem is distinct: NN contributes to the torque estimator used to infer $G$, and therefore must be propagated within the GUM framework~\cite{GUM}.
At the same time, symmetry-based suppression is not ``free'': if both the desired source-mass signal and the environmental field couple through the same quadrupole channel, reducing that channel suppresses both together.
Accordingly, the perfectly symmetric cross analysed below is not presented as a complete $G$-measurement geometry, but as an idealized rejection benchmark that clarifies the long-wavelength limit.
A practical experiment must either retain a controlled residual quadrupole (engineered anisotropy) or use a source-mass scheme that couples to a different spatial channel.

We follow the measurement-model approach of the GUM: NN is treated as a standard-uncertainty contribution within an explicit measurement model---namely, (i) a measurement equation is stated, (ii) the relevant input quantities and their standard uncertainties are identified, and (iii) uncertainties are propagated to $u(\hat{G})$ using the GUM law of propagation of uncertainty (including correlations where relevant).
Here, ``propagation'' refers to uncertainty propagation through the measurement equation (not propagation of physical fields in space or time).
A key technical step beyond a purely algebraic GUM treatment is an explicit evaluation of the torque-estimator standard uncertainty $u(\hat{\tau})$ for correlated noise via the estimator transfer function, together with a clear separation between stationary environmental inputs and non-stationary baseline drift.
Relative to the interferometric Newtonian-noise literature, the novelty here is therefore not a new atmospheric field model but its adaptation to torsion-balance $G$ metrology: estimator-dependent uncertainty propagation in a GUM-consistent framework, benchmark transfer functions for low-order spatial coupling, and an explicit metrological separation between stationary inputs and non-stationary drift.

Accordingly, this paper makes five contributions:
\begin{enumerate}
\item[(i)] an estimator-dependent evaluation of NN contributions to $u(\hat{G})$ within the GUM framework;
\item[(ii)] analytic transfer functions from spatio-temporal surface-density fluctuations to torque, including baseline suppression;
\item[(iii)] a closed-form extension from the dumbbell to an idealized symmetric cross configuration, quantifying the limiting case of quadrupole cancellation and the resulting signal--noise trade-off;
\item[(iv)] a clear separation between stationary correlated environmental inputs, for which PSD-based propagation applies, and non-stationary baseline wandering, which requires separate metrological treatment; and
\item[(v)] a realistic atmospheric-pressure benchmark based on the global infrasound pressure-noise envelope (\emph{pressure} NLNM/NHNM)~\cite{Marty2021}.
\end{enumerate}
The central claim is therefore modest but practically important: in the benchmark range treated here, atmospheric background NN is not large enough to account for the present scatter in published $G$ values, yet as experiments push to lower systematic floors it is approaching a regime where it should be explicitly bounded, monitored, and possibly subtracted.
An experiment-by-experiment assessment, however, requires apparatus-specific transfer functions and estimators.

\section{Measurement model and uncertainty propagation}
\subsection{Measurement equation}
We use the standard linear measurement equation
\begin{equation}
\hat{G} = \frac{\hat{\tau}}{C_G},
\label{eq:Ghat}
\end{equation}
where $\hat{\tau}$ is the estimated gravitational signal torque and $C_G$ is a geometrical response coefficient determined from the experiment's field model and calibration.

\subsection{GUM propagation and the environmental term}
With inputs $X=(\hat{\tau},C_G)$, the first-order GUM uncertainty is
\begin{equation}
u^2(\hat{G})\simeq \left(\frac{\partial G}{\partial \tau}\right)^2 u^2(\hat{\tau})+
\left(\frac{\partial G}{\partial C_G}\right)^2 u^2(C_G)+
2\frac{\partial G}{\partial \tau}\frac{\partial G}{\partial C_G}u(\hat{\tau},C_G),
\label{eq:GUMprop}
\end{equation}
with sensitivity coefficients $\partial G/\partial\tau=1/C_G$ and $\partial G/\partial C_G=-\hat{\tau}/C_G^2$.
We focus on the NN contribution to the torque estimator uncertainty, $u_{\rm env}(\hat{\tau})$.
Its contribution to the relative standard uncertainty is
\begin{equation}
 u_{r,{\rm env}}(G)\equiv \frac{u_{\rm env}(\hat{G})}{|\hat{G}_{\rm sig}|}
 \simeq \frac{u_{\rm env}(\hat{\tau})}{|\hat{\tau}_{\rm sig}|},
\qquad \hat{\tau}_{\rm sig}\equiv\langle\hat{\tau}\rangle,
\label{eq:ur_env}
\end{equation}
where $\hat{G}_{\rm sig}=\hat{\tau}_{\rm sig}/C_G$.
We normalize by the signal estimate rather than by the instantaneous total estimate because the purpose of $u_{r,{\rm env}}(G)$ is to quantify environmental contamination relative to the intended $G$-bearing signal channel.

\paragraph{Remarks on covariance and geometric uncertainty.}
In Eq.~\eqref{eq:GUMprop} the covariance term accounts for statistical dependence between $\hat{\tau}$ and $C_G$.
For the NN contribution analysed here, it is natural to treat $C_G$ as determined independently of the stochastic NN time series that perturbs $\hat{\tau}$ during a measurement run, so that $u(\hat{\tau},C_G)\approx 0$ for this pathway.
If an experiment exhibits shared environmental dependencies (e.g. temperature fluctuations affecting both the torque estimate and the geometric/calibration factor), the covariance term can be retained without modifying the remainder of the framework.

\section{Estimator-dependent variance for correlated torque noise}
Let the measured torque time series be $\tau(t)=\tau_{\rm sig}(t)+n(t)$ with zero-mean stationary environmental torque noise $n(t)$.
This section addresses only the stationary component that can be described by a PSD; non-stationary baseline wandering is separated explicitly in Sec.~\ref{sec:nonstationary}.
A general linear estimator can be written as
\begin{equation}
\hat{\tau}=\int_{-\infty}^{\infty}w(t)\,\tau(t)\,{\rm d}t,
\label{eq:estimator}
\end{equation}
where the (chosen) weight function $w(t)$ encodes averaging, demodulation, regression subtraction, etc.
We treat $w(t)$ as deterministic because it represents the specified estimation procedure; the stochasticity resides in the measured process $\tau(t)$.

For a one-sided torque PSD $S_\tau(\omega)$ defined for $\omega\ge 0$, the estimator variance is
\begin{equation}
 u^2(\hat{\tau})=\int_{0}^{\infty}\frac{{\rm d}\omega}{2\pi}\,|W(\omega)|^2\,S_\tau(\omega),
\label{eq:var_general}
\end{equation}
with $W(\omega)$ the Fourier transform of $w(t)$.\footnote{Equivalently, one may use a two-sided PSD and integrate over $\omega\in(-\infty,\infty)$; for a real process the two conventions differ by a factor of two for $\omega>0$.}

\paragraph{Spectral-density conventions.}
We use the term \emph{power spectral density} (PSD) for spectra such as $S_\tau(\omega)$, with SI units of (N\,m)$^2$/Hz for torque.
When presenting spectra in figures, we also use the corresponding \emph{amplitude spectral density} (ASD). For the torque channel,
\begin{equation}
 \mathrm{ASD}_\tau(\omega)\equiv\sqrt{S_\tau(\omega)},
\end{equation}
and analogously for other channels such as pressure or surface density.
Thus $\mathrm{ASD}_\tau$ has SI units of \si{\newton \, \metre \, \hertz^{-1/2}}.

\subsection{Stationary OU benchmark for environmental-input averaging}
\label{sec:OUbenchmark}
\begin{figure}[t]
\centering
\includegraphics[width=0.72\linewidth]{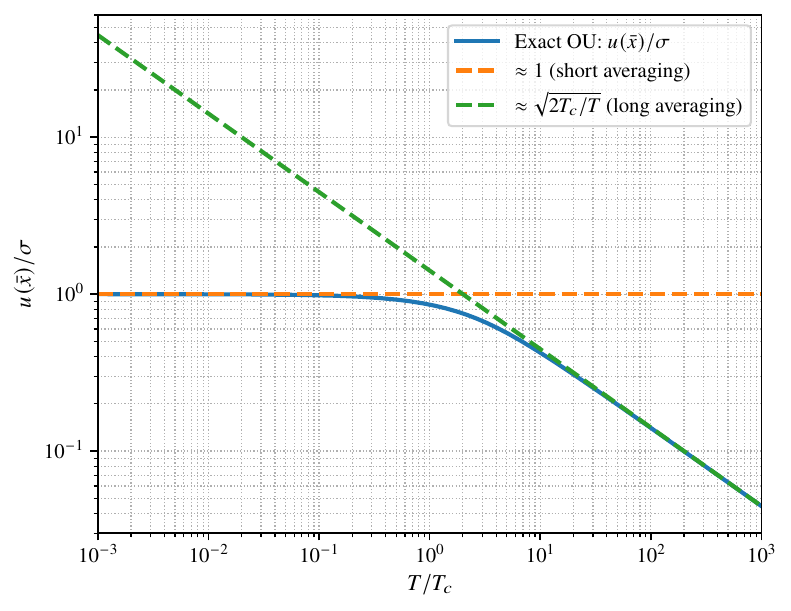}
\caption{Normalized standard uncertainty of the sample mean for the stationary OU benchmark, Eq.~\eqref{eq:OUmean}.}
\label{fig:OU}
\end{figure}

To make the role of finite correlation time explicit for the stationary part of the environmental input, we use the Ornstein--Uhlenbeck (OU) model as a minimal benchmark.
Here $x(t)$ denotes a stationary stochastic input with variance $\sigma^2$ and correlation time $T_c$.
In the present context, $x(t)$ should be read as the environmental torque noise $n(t)$ itself, or as a single stationary component of the environmental field after linear coupling to torque; it is \emph{not} a model of the absolute equilibrium angle of the torsion balance.
For an OU process $x(t)$, the standard uncertainty of the sample mean over duration $T$ is
\begin{equation}
 u^2(\bar{x})=\frac{2\sigma^2 T_c}{T^2}\left[T-T_c\left(1-e^{-T/T_c}\right)\right].
\label{eq:OUmean}
\end{equation}
Figure~\ref{fig:OU} shows the normalized reduction $u(\bar{x})/\sigma$.
For $T\gg T_c$, Eq.~\eqref{eq:OUmean} reduces to the expected $T^{-1/2}$ averaging law; for $T\lesssim T_c$, averaging becomes inefficient and $u(\bar{x})$ approaches $\sigma$.
This saturation is useful as a benchmark for long-correlation \emph{stationary} noise, but it should not be confused with a true random walk or Brownian drift, which is non-stationary and has no finite equilibrium variance.

\subsection{Non-stationary drift as a separate metrological contribution}
\label{sec:nonstationary}
Many Cavendish-type measurements are also affected by slow baseline wandering of the balance equilibrium, source-mass configuration, laboratory environment, or other instrumental states.
Such processes need not have a mean-reverting timescale within a single run and therefore fall outside the stationary OU benchmark above.
They should be modeled separately (for example by random-walk, $1/f^\alpha$, or trend models, depending on the apparatus) or, if insufficiently characterized, treated conservatively as Type-B components or as conservative Type-A bounds within the GUM framework.
The PSD-based propagation developed below therefore applies to the stationary part of the environmental NN input, while non-stationary drift remains a distinct problem in $G$ metrology rather than something solved by the OU benchmark itself.

\section{Thermal torque noise reference}
Thermal torque noise provides a convenient reference floor for torsion balances.
For viscous (dashpot) damping, the fluctuation--dissipation theorem gives a white (one-sided) thermal torque PSD
\begin{equation}
S_{\tau,{\rm th}}(\omega)=4k_{\rm B}T\,\gamma_{\rm d}
= \frac{4k_{\rm B}T\,\kappa_{\rm eff}}{Q\,\omega_0},
\qquad Q \equiv \frac{\kappa_{\rm eff}}{\gamma_{\rm d}\,\omega_0},
\label{eq:Sth}
\end{equation}
where $\gamma_{\rm d}$ denotes the viscous rotational damping coefficient. This avoids a collision with the gravity-gradient symbol $\Gamma_{ij}$ used below. The expression is valid in the off-resonant (estimator) band $\omega\ll \omega_0$.

For a simple DC mean over duration $T_{\rm th}$, $w(t)=1/T_{\rm th}$ for $t\in[0,T_{\rm th}]$ and
\begin{equation}
W(\omega)=\frac{1}{T_{\rm th}}\int_0^{T_{\rm th}} e^{i\omega t}\,dt
= e^{i\omega T_{\rm th}/2}\,\mathrm{sinc}\!\left(\frac{\omega T_{\rm th}}{2}\right),
\end{equation}
so that $|W(0)|=1$.
If the torque PSD is effectively flat across the estimator band, then inserting $W(\omega)$ into Eq.~\eqref{eq:var_general} yields
\begin{equation}
 u_{\rm th}(\hat{\tau})\simeq \sqrt{\frac{S_{\tau,{\rm th}}}{2T_{\rm th}}} \propto T_{\rm th}^{-1/2}.
\label{eq:uth_white}
\end{equation}
Under the one-sided PSD convention, the factor $1/(2T_{\rm th})$ follows from
$\int_0^{\infty}(d\omega/2\pi)\,\mathrm{sinc}^2(\omega T_{\rm th}/2)=1/(2T_{\rm th})$.

A structural-damping (loss-angle) model yields a red low-frequency spectrum $S_{\tau,{\rm th}}\propto 1/\omega$; in metrological use, the measured off-resonance spectrum should anchor the effective dissipation model in the relevant estimator band.

\section{From gravity gradients to torque}
\subsection{General coupling and an ``equivalent gradient''}
Let $\Gamma_{ij}=\partial g_j/\partial x_i$ be the gravity-gradient tensor at the pendulum.
For torsion about $z$ and a baseline oriented along $x$, the dominant low-order coupling is typically to $\Gamma_{xy}=\partial g_y/\partial x$.
To linear order in gradients, the torque may be written as
\begin{equation}
\tau_z(t)=C_\Gamma\,\Gamma_{xy}(t),
\label{eq:tau_CG}
\end{equation}
where $C_\Gamma$ is a geometry factor determined by the pendulum's mass distribution (it is proportional to the relevant quadrupole moment).

Equation~\eqref{eq:tau_CG} motivates an \emph{equivalent gradient} representation: any torque noise level $u(\hat{\tau})$ can be translated into an equivalent gradient uncertainty
\begin{equation}
\sigma_{\Gamma,{\rm eq}}=\frac{u(\hat{\tau})}{|C_\Gamma|},
\label{eq:sigmaGammaEq}
\end{equation}
which is convenient for comparing different torsion balances independently of $C_G$. Throughout, equivalent gradients are reported in SI units of \si{\per\square\second}.

\subsection{Dumbbell as a transparent worst-case model}
For a two-mass dumbbell with point masses $m$ at $(\pm l,0)$,
$C_\Gamma=2ml^2$ and $\tau_z=2ml^2\,\Gamma_{xy}$.
The dumbbell is not intended as a realistic model of quadrupole-suppressed pendula; rather, it serves as a transparent \emph{upper bound} on NN coupling when the pendulum quadrupole moment is not minimized.

\subsection{Cross configuration as an idealized rejection benchmark}
For a symmetric cross with four masses $m$ at $(\pm l,0)$ and $(0,\pm l)$, the net coupling to a uniform $\Gamma_{xy}$ cancels by symmetry.
The leading environmental coupling then arises from higher spatial derivatives of the gravity field (schematically, gradients of $\Gamma$) and is therefore more strongly suppressed at long wavelength.
This exactly symmetric cross is best interpreted as an idealized rejection benchmark, not as a complete measurement geometry for $G$.
If the desired source-mass signal also enters through the same quadrupole ($l=2$) channel, perfect cancellation would suppress the signal together with the environmental NN.
A practical instrument must therefore retain a controlled residual quadrupole (``engineered anisotropy'') or adopt a source-mass scheme that couples to a higher spatial channel.

To parameterize this trade-off phenomenologically, it is convenient to introduce a residual quadrupole fraction
\begin{equation}
C_\Gamma(\epsilon)=\epsilon\,C_\Gamma^{\rm(db)},\qquad 0\le \epsilon \le 1,
\label{eq:epsilon_tradeoff}
\end{equation}
where $\epsilon=0$ denotes the perfectly symmetric cross and $\epsilon=1$ a dumbbell-like reference.
For signal and environmental terms that enter through the same uniform-$\Gamma_{xy}$ channel, both $\tau_{\rm sig}$ and $u_{\rm env}(\hat{\tau})$ then scale approximately with $\epsilon$.
As emphasized later by Eq.~\eqref{eq:ur_env_grad}, symmetry alone does not automatically reduce the relative uncertainty in $G$; the practical advantage of multipole engineering lies instead in selective rejection of environmental fields relative to the source field, or in shifting the measurement to a different spatial channel.
In Sec.~\ref{sec:baseline} we quantify the idealized $\epsilon=0$ limit at the transfer-function level via the baseline factor $A_\times(kl)$.

\section{Newtonian noise from surface-density fluctuations}
\subsection{Surface-density model}
\label{sec:surfmodel}
We model environmental mass perturbations by an effective surface density field $\delta\Sigma(\mathbf{r},t)$ on the plane $z=0$.
This approximation captures, for example, column air-mass changes associated with pressure fluctuations and shallow hydrology, and it is widely used in NN modelling~\cite{Harms2015,Harms2019,Creighton2008}.

\paragraph{Fourier convention.}
We use the two-dimensional spatial Fourier transform
\begin{equation}
\delta\Sigma(\mathbf{r},t)=\int\frac{{\rm d}^2\mathbf{k}}{(2\pi)^2}\,\delta\Sigma(\mathbf{k},t)\,e^{i\mathbf{k}\cdot\mathbf{r}},\qquad
\delta\Sigma(\mathbf{k},t)=\int {\rm d}^2\mathbf{r}\,\delta\Sigma(\mathbf{r},t)\,e^{-i\mathbf{k}\cdot\mathbf{r}}.
\label{eq:FTconv}
\end{equation}

With this convention, the Newtonian potential perturbation at height $z_0$ above the plane is
\begin{equation}
\delta\Phi(\mathbf{r},z_0,t)=-\int\frac{{\rm d}^2\mathbf{k}}{(2\pi)^2}\,\frac{2\pi G}{k}\,e^{-kz_0}\,\delta\Sigma(\mathbf{k},t)\,e^{i\mathbf{k}\cdot\mathbf{r}}.
\label{eq:Phi_realspace}
\end{equation}
For a single spatial Fourier mode $\mathbf{k}$, the corresponding mode coefficient is
$\delta\Phi(\mathbf{k},z_0,t)=-(2\pi G/k)e^{-kz_0}\,\delta\Sigma(\mathbf{k},t)$.
The shear-gradient coefficient for that mode is therefore
\begin{equation}
\Gamma_{xy}(\mathbf{k},t)=-2\pi G\,e^{-kz_0}\,\frac{k_xk_y}{k}\,\delta\Sigma(\mathbf{k},t),
\label{eq:Gamma_k}
\end{equation}
and the physical field $\Gamma_{xy}(\mathbf{r},t)$ is obtained by inserting Eq.~\eqref{eq:Gamma_k} into the inverse transform~\eqref{eq:FTconv}.
In what follows, expressions such as $\Gamma_{xy}(\mathbf{k},t)$ and $\tau_z(\mathbf{k},t)$ refer to these mode coefficients. Only variances and PSDs are used later, so this sign convention does not affect the benchmark uncertainty levels.

\paragraph{Limitations of the surface-density approximation.}
A full atmospheric model would treat a three-dimensional density perturbation $\delta\rho(\mathbf{r},z,t)$.
In that case, the plane-wave kernel in Eq.~\eqref{eq:Phi_realspace} weights altitude by $e^{-kz}$, so the relevant quantity is the vertically weighted column
\begin{equation}
\tilde{\Sigma}(\mathbf{k},t)\equiv \int_0^{\infty}{\rm d}z\,\delta\rho(\mathbf{k},z,t)\,e^{-kz}.
\end{equation}
Replacing $\tilde{\Sigma}$ by an unweighted column mass (or by $\delta p/g$) is therefore generally conservative because $e^{-kz}\le 1$.
For an exponential vertical profile $\delta\rho\propto e^{-z/H}$ one finds $\tilde{\Sigma}=\Sigma/(1+kH)$, i.e.\ the surface-density approximation overestimates the coupling by a factor $1+kH$.
For illustration, $H=\SI{8}{km}$ gives $\tilde{\Sigma}/\Sigma\approx 1/81$ at horizontal scale $k^{-1}=\SI{100}{m}$ and $\approx 1/801$ at $k^{-1}=\SI{10}{m}$.
In addition, at $z_0\sim\SI{1}{m}$ the mapping from pressure to column mass is only approximate: non-hydrostatic fluctuations, boundary-layer turbulence, and building-specific pressure patterns can modify the relation between a local barometer record and the gravity-gradient field.
Because the Marty pressure NLNM/NHNM are outdoor background envelopes, indoor laboratory pressure fields can also be larger owing to building acoustics, ventilation, and room-scale flow structure.
The atmospheric benchmark in Sec.~\ref{sec:atmos} should therefore be interpreted as an order-of-magnitude envelope and a conservative design reference in the absence of site-specific array measurements.

\subsection{Transfer functions and baseline factors}
\label{sec:baseline}
For a dumbbell, the torque from a single Fourier mode is
\begin{equation}
\tau_z(\mathbf{k},t)=-4\pi G\,m\,l\,e^{-kz_0}\,\frac{k_y}{k}\,\sin(k_x l)\,\delta\Sigma(\mathbf{k},t).
\label{eq:tau_dumbbell_k}
\end{equation}
Writing $k_x=k\cos\varphi$ and $k_y=k\sin\varphi$, the squared coupling contains the direction-dependent factor
$\sin^2\varphi\,\sin^2(kl\cos\varphi)$.
Assuming isotropy in $\mathbf{k}$, it is convenient to separate this purely geometric reduction into the \emph{baseline factor}
\begin{equation}
A(kl)=\frac{1}{2\pi}\int_0^{2\pi}\sin^2\varphi\,\sin^2\!\left(kl\cos\varphi\right)\,{\rm d}\varphi
=\frac14\left[1-J_0(2kl)-J_2(2kl)\right].
\label{eq:A_dumbbell}
\end{equation}
where $J_n$ denotes the Bessel function of the first kind of order $n$.
The baseline factor quantifies long-wavelength cancellation ($kl\ll1$) and approaches an $\mathcal{O}(1)$ constant for $kl\gg1$.
The torque PSD becomes
\begin{equation}
S_\tau(\omega)=(4\pi G m l)^2\int_0^\infty\frac{k\,{\rm d}k}{2\pi}\,e^{-2kz_0}\,A(kl)\,P_\Sigma(k,\omega),
\label{eq:Stau_general}
\end{equation}
where $P_\Sigma(k,\omega)$ is the spatio-temporal spectrum of $\delta\Sigma$.
Equation~\eqref{eq:Stau_general} is the master expression used below. Section~\ref{sec:atmos} closes it by an explicit $k$--$\omega$ mapping, whereas Sec.~7.1 uses a separable ansatz; these are alternative benchmark models, not simultaneous assumptions.

For the symmetric cross, the single-mode torque is
\begin{equation}
\tau_{z,\times}(\mathbf{k},t)=4\pi G\,m\,l\,e^{-kz_0}\,
\left[\frac{k_x}{k}\sin(k_y l)-\frac{k_y}{k}\sin(k_x l)\right]\delta\Sigma(\mathbf{k},t),
\label{eq:tau_cross_k}
\end{equation}
and the corresponding baseline factor is
\begin{equation}
A_\times(kl)=\frac{1}{2\pi}\int_0^{2\pi}
\left[\cos\varphi\,\sin(kl\sin\varphi)-\sin\varphi\,\sin(kl\cos\varphi)\right]^2{\rm d}\varphi.
\label{eq:A_cross_def}
\end{equation}
For $kl\ll1$,
\begin{equation}
A(kl)\simeq \frac{(kl)^2}{8},\qquad
A_\times(kl)\simeq \frac{(kl)^6}{1152},
\label{eq:smallk}
\end{equation}
making explicit the additional $(kl)^4$ suppression in power produced by quadrupole cancellation (see Appendices~A--B for expansions).
Figure~\ref{fig:baseline} compares $A$ and $A_\times$.

\begin{figure}[t]
\centering
\includegraphics[width=0.78\linewidth]{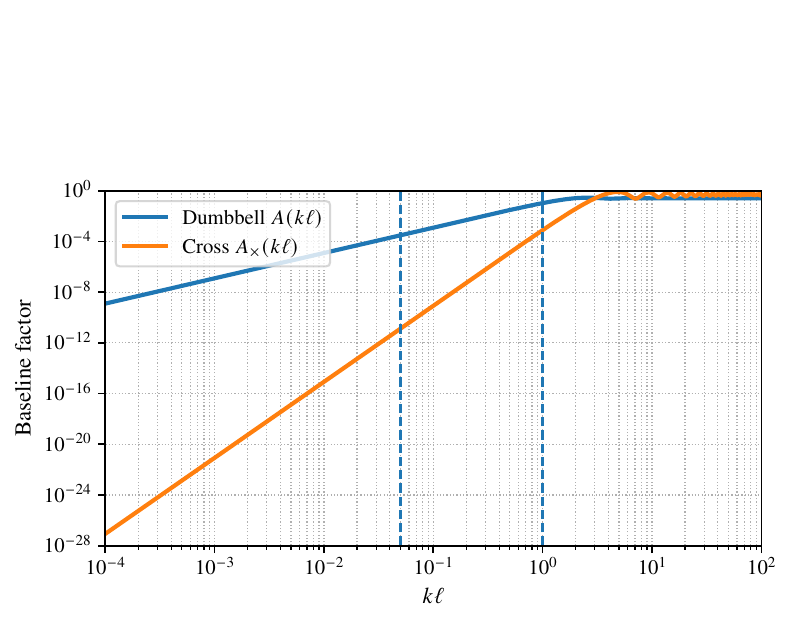}
\caption{Baseline suppression factors for isotropic surface-density fields: dumbbell $A(kl)$, Eq.~\eqref{eq:A_dumbbell}, and symmetric cross $A_\times(kl)$, Eq.~\eqref{eq:A_cross_def}. Dashed vertical lines indicate $k\sim 1/z_0$ and $k\sim 1/l$ for representative laboratory scales $l=\SI{5}{cm}$, $z_0=\SI{1}{m}$.}
\label{fig:baseline}
\end{figure}

The dumbbell and symmetric-cross factors should therefore be read as two limiting cases.
A practical engineered-anisotropy design will lie between them, retaining finite source-mass coupling while partially suppressing long-wavelength environmental gradients.
Deriving that interpolation requires the detailed mass distribution and source modulation of a specific experiment and is left for future work.

\section{Spatio-temporal environmental inputs}
The central experimental input is the spatio-temporal spectrum $P_\Sigma(k,\omega)$.
Single-point measurements constrain only the temporal PSD; arrays are required to characterize spatial structure and hence the $k$-dependence that controls baseline suppression.
We consider two complementary but logically separate closures of Eq.~\eqref{eq:Stau_general}: a simple separable model useful for design studies, and a data-driven atmospheric benchmark that maps temporal frequency to horizontal wavenumber through an assumed phase velocity. These are alternative benchmarks, not simultaneous assumptions about one exact joint spectrum.

\subsection{Separable stationary correlation model (exponential $\times$ OU)}
As a tractable benchmark, assume separable spatial and temporal correlations,
\begin{equation}
\langle \delta\Sigma(\mathbf{r},t)\,\delta\Sigma(\mathbf{0},0)\rangle=\sigma_\Sigma^2\,e^{-r/L}\,e^{-|t|/T_c},
\label{eq:sep_corr}
\end{equation}
with RMS surface-density fluctuation $\sigma_\Sigma$, correlation length $L$, and correlation time $T_c$.
The corresponding spatial spectrum is
\begin{equation}
P_\Sigma(k)=\frac{2\pi\sigma_\Sigma^2L^2}{(1+k^2L^2)^{3/2}},
\label{eq:P_k}
\end{equation}
and the OU temporal PSD is
\begin{equation}
S(\omega)=\frac{2T_c}{1+\omega^2T_c^2}.
\label{eq:OU_PSD}
\end{equation}
The full spatio-temporal spectrum is $P_\Sigma(k,\omega)=P_\Sigma(k)\,S(\omega)$.
In this subsection, space and time are therefore assumed separable by construction.
This model makes explicit how $L$ controls baseline suppression and how $T_c$ controls averaging, and it is therefore useful for experimental design and sensor-placement studies even when the true spectrum is more complex.
Because the temporal factor is OU, however, this subsection addresses only the stationary part of the environmental field and is not intended as a model of non-stationary apparatus drift.
The exponential spatial correlation is a standard parametric choice in spatial statistics~\cite{Cressie1993}.

\subsection{Data-driven atmospheric benchmark: pressure NLNM/NHNM}
\label{sec:atmos}
To connect the framework to realistic environmental inputs, we use the global low- and high-noise models for ambient pressure spectra compiled from the infrasound network~\cite{Marty2021}.
Following Marty \emph{et al.}\ we refer to these as the \emph{pressure} NLNM/NHNM to avoid confusion with the Peterson seismic NLNM/NHNM models for ground motion~\cite{Peterson1993}.

For pressure fluctuations, an effective surface-density conversion follows from hydrostatic balance $p=g\Sigma$:
\begin{equation}
\delta\Sigma\simeq \frac{\delta p}{g},\qquad S_\Sigma(f)\simeq \frac{S_p(f)}{g^2},
\label{eq:Sp_to_Ssigma}
\end{equation}
where $g$ is the local standard acceleration of gravity (\SI{9.81}{\metre\per\square\second}).
Here $\Sigma$ has SI units of \si{\kilogram\per\square\metre} (column mass per unit area).

A full NN calculation requires the joint dependence on $(k,\omega)$; arrays are required to measure this directly.
For a design-level benchmark we bracket the mapping from temporal frequency to horizontal wavenumber using a phase-velocity relation
\begin{equation}
k(f)=\frac{2\pi f}{v_{\rm ph}},
\label{eq:komega}
\end{equation}
with $v_{\rm ph}=\SI{340}{\metre\per\second}$ (acoustic) and $v_{\rm ph}\sim\SI{10}{\metre\per\second}$ (slow advection).
The slow-advective value is used only as an order-of-magnitude benchmark for slowly transported pressure patterns (for example background wind or building-scale advection/ventilation); it is not presented as a universal measured phase velocity for all laboratories.
For the advective interpretation, Eq.~\eqref{eq:komega} corresponds to a ``frozen field'' (Taylor hypothesis) mapping between temporal and spatial spectra~\cite{Taylor1938}.
In this benchmark, the isotropic baseline factor $A(kl)$ is evaluated in closed form (Eq.~\eqref{eq:A_dumbbell}).

\begin{figure}[t]
\centering
\includegraphics[width=0.90\linewidth]{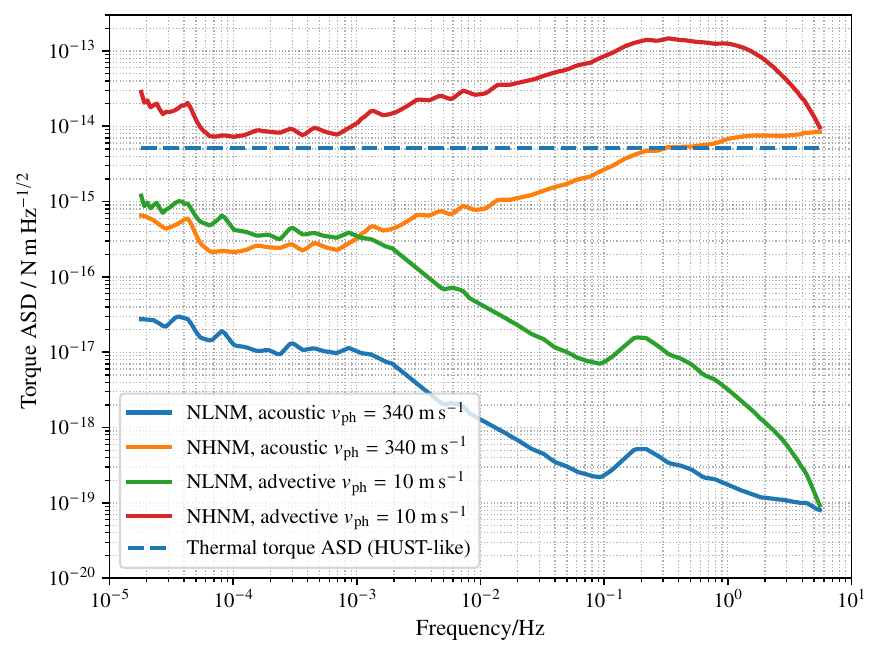}
\caption{Atmospheric pressure benchmark propagated to torque for a representative dumbbell benchmark ($l=\SI{5}{cm}$, $m=\SI{0.53}{kg}$, $z_0=\SI{1}{m}$).
Pressure PSD envelopes (pressure NLNM/NHNM) are taken from~\cite{Marty2021} and converted using Eq.~\eqref{eq:Sp_to_Ssigma}.
The torque ASDs are then computed via Eq.~\eqref{eq:tauASD_benchmark}, with the mapping from $f$ to $k$ bracketed by Eq.~\eqref{eq:komega}.
The benchmark is illustrative and is not intended as an apparatus-specific reconstruction of HUST-99, UWash-00, or the BIPM balances; its role is to show how atmospheric pressure spectra translate into torque for a transparent low-order coupling model.
The dashed line is the thermal torque ASD obtained from Eq.~\eqref{eq:Sth} (as $\mathrm{ASD}_\tau=\sqrt{S_{\tau,\mathrm{th}}}$), using the benchmark plotting parameters $T=\SI{300}{K}$, $\kappa_{\rm eff}=2.74\times10^{-9}$~\si{\newton\metre\per\radian}, $Q=3433$, and $f_0=7.77\times10^{-5}$~\si{Hz}.
}
\label{fig:pressure_to_torque}
\end{figure}

For the plots we express spectra in terms of amplitude spectral densities.
Writing $\mathrm{ASD}_p(f)\equiv\sqrt{S_p(f)}$ and using Eq.~\eqref{eq:Sp_to_Ssigma}, the corresponding surface-density ASD is $\mathrm{ASD}_\Sigma(f)=\mathrm{ASD}_p(f)/g$.
Under the plane-wave benchmark mapping $k=k(f)$, the torque ASD shown in Fig.~\ref{fig:pressure_to_torque} is approximated by
\begin{equation}
 \mathrm{ASD}_\tau(f)\simeq (4\pi G m l)\,e^{-k(f)z_0}\,\sqrt{A(k(f)l)}\,\mathrm{ASD}_\Sigma(f).
\label{eq:tauASD_benchmark}
\end{equation}
Here $e^{-k(f)z_0}\sqrt{A(k(f)l)}$ acts as a deterministic frequency-dependent transfer factor after the benchmark closure $k=k(f)$ has been imposed.
Equation~\eqref{eq:tauASD_benchmark} is therefore not the Fourier transform of a time-domain product and does not introduce a convolution; it is a one-dimensional spectral benchmark obtained after closing the joint $(k,\omega)$ dependence by Eq.~\eqref{eq:komega}.
Figure~\ref{fig:pressure_to_torque} shows the resulting torque amplitude spectral densities for a representative dumbbell and compares them to the thermal torque ASD.

\subsection{Seismic Newtonian noise and ground motion}
While Sec.~\ref{sec:atmos} focuses on atmospheric pressure as a concrete benchmark, gravity-gradient fluctuations can also be driven by seismic waves and local ground deformation.
In many terrestrial precision experiments, especially at low frequencies, seismic NN can be comparable to or larger than atmospheric NN; the relative importance is strongly site dependent (geology, depth, wave content) and depends on the estimator band~\cite{Harms2015,Harms2019}.
In seismology, the acronyms NLNM/NHNM are widely used for the Peterson reference models of ground-motion spectra~\cite{Peterson1993}.
Combined with transfer functions from ground displacement fields to gravity gradients, these provide an analogous ``background envelope'' for seismic NN.
A full seismic NN treatment is beyond our present scope, but the estimator-based propagation formalism of Secs.~2--3 and the baseline suppression factors of Sec.~\ref{sec:baseline} apply unchanged once site-specific spatio-temporal spectra (or array-derived cross-spectra) of $\Gamma_{xy}$ are specified.

\section{Representative experimental contexts}
The generic transfer functions derived in Secs.~2--6 are intended as benchmark couplings, not as faithful re-analyses of each historical $G$ apparatus.
Historical torsion-balance experiments used materially different pendulum and source-mass layouts (beam/dumbbell, rectangular plate, and four-test-mass torsion-strip balances) as well as different estimators and operating modes.
Table~\ref{tab:expcontext} therefore lists only directly reported or review-summarized contextual information.
UWash-00 employed a flat rectangular plate and should be analysed with a plate-specific multipole treatment, not with the dumbbell baseline factor $A(kl)$ derived in Sec.~\ref{sec:baseline}.
Likewise, the BIPM balances used four test masses on a torsion-strip balance and require an apparatus-specific transfer function.
The dumbbell and symmetric-cross models developed here should therefore be interpreted as benchmark envelopes for low-order spatial coupling, not as literal reconstructions of every historical geometry.
A claim about whether atmospheric NN dominates any one historical apparatus would require a dedicated apparatus-specific re-analysis.

\begin{table}[H]
\centering
\footnotesize
\setlength{\tabcolsep}{4pt}
\renewcommand{\arraystretch}{1.15}
\caption{Representative torsion-balance $G$ experiments used for contextual comparison.
Only source-traceable literature quantities are listed.
We intentionally do not tabulate experiment-specific benchmark couplings such as $C_\Gamma$ or $\kappa_{\rm eff}$, because these require a dedicated apparatus re-analysis and are not uniquely extractable from the short experimental papers alone.
For consistency with common CODATA-style labels, we retain the label BIPM-14 for the improved BIPM result although the principal short paper appeared in 2013.}
\label{tab:expcontext}
\begin{adjustbox}{max width=\linewidth}
\begin{tabular}{p{1.6cm}p{4.2cm}p{1.7cm}p{1.4cm}p{1.6cm}p{4.3cm}}
\toprule
Experiment & Reported geometry / operating mode & Reported period & Reported $Q$ & Published $u_r(G)$ & Implication for the present model\\
\midrule
HUST time-of-swing case\cite{HUSTReview2014} & Time-of-swing HUST configuration as summarized in the review literature; exact pendulum and source-mass details are not unpacked here because several HUST campaigns are discussed in that review. & \SI{3484}{s} (review summary) & $3.6\times10^{4}$ (review summary) & 105 ppm & Contextual example only; a source-traceable transfer function would require the original apparatus paper and full geometry.\\
UWash-00\cite{UWash00} & Flat rectangular plate pendulum with angular-acceleration feedback and rotating attractor spheres. & --- & --- & 14 ppm & Requires a plate-specific multipole / transfer-function treatment; the dumbbell baseline factor $A(kl)$ is not the appropriate apparatus model.\\
BIPM-01\cite{BIPM01} & Torsion-strip balance operated in electrostatic servo and free-deflection (Cavendish) modes. & --- & --- & 41 ppm & Fourfold source/test-mass geometry; not reduced here to a single benchmark $C_\Gamma$ or dumbbell/cross baseline.\\
BIPM-14\cite{BIPM13} & Rebuilt torsion-strip balance with four test masses and four source masses; electrostatic servo and free-deflection modes. & \SI{120}{s} & $\sim 10^{5}$ & 27 ppm & Requires a dedicated four-mass transfer function; the actual signal channel is not identical to the uniform-$\Gamma_{xy}$ quadrupole benchmark.\\
\bottomrule
\end{tabular}
\end{adjustbox}
\end{table}

\section{When does Newtonian noise matter for $u_r(G)$? Present scale and relative targets below $10^{-6}$}
A Newtonian-noise (NN) contribution becomes metrologically relevant only if it produces a non-negligible contribution to the relative standard uncertainty $u_r(G)$.
This depends on the size of the gravitational signal gradient and on how the environmental gradients are filtered by the chosen estimator (averaging, demodulation, regression), not simply on whether the NN torque exceeds the thermal torque floor.
To make this dependence explicit, it is useful to express both the signal and the environmental contribution in terms of equivalent gravity gradients.

\subsection{From gravity gradients to a relative uncertainty scale}
In the gradient limit, both the signal torque and the environmental contribution can be expressed in terms of an equivalent gravity gradient.
Define the signal-equivalent gradient by
\begin{equation}
\Gamma_{\rm sig} \equiv \frac{\tau_{\rm sig}}{C_\Gamma},
\label{eq:Gamma_sig_def}
\end{equation}
and let $\sigma_\Gamma$ denote the estimator-standard contribution of environmental gradients (after applying the chosen estimator, including averaging, demodulation, and/or regression).
Then, using Eqs.~\eqref{eq:ur_env} and \eqref{eq:tau_CG},
\begin{equation}
 u_{r,{\rm env}}(G)
 \simeq \frac{|C_\Gamma|\,\sigma_\Gamma}{|\tau_{\rm sig}|}
 =\frac{\sigma_\Gamma}{\Gamma_{\rm sig}}.
\label{eq:ur_env_grad}
\end{equation}
Equation~\eqref{eq:ur_env_grad} makes the key point explicit: at leading order, the relevance of environmental NN depends on the ratio of \emph{environmental} to \emph{signal} gradients and is largely independent of the pendulum geometry factor $C_\Gamma$.

A convenient heuristic for the signal scale is $\Gamma_{\rm sig}\sim GM/r^3$, where $M$ is a characteristic attractor mass and $r$ a characteristic separation.

\label{sec:benchmark_gamma}
To avoid giving the impression of an apparatus-specific reconstruction, we use two rounded benchmark signal gradients in the discussion below,
\begin{equation}
\Gamma_{\rm sig}^{\rm (bench)}\in\left\{10^{-7},\;10^{-6}\right\}\ \si{\per\square\second}.
\label{eq:Gamma_sig_bench}
\end{equation}
These numbers are introduced as internal benchmarks, not as literature values extracted from any single experiment.
They are motivated by the heuristic $GM/r^3$: for example,
$M=\SI{1}{kg}$ at $r=\SI{10}{cm}$ gives $\Gamma\approx 6.7\times10^{-8}$~\si{\per\square\second}, while $M=\SI{10}{kg}$ at $r=\SI{5}{cm}$ gives $\Gamma\approx 5.3\times10^{-6}$~\si{\per\square\second}.
We therefore use $10^{-7}$ and $10^{-6}$~\si{\per\square\second} as two conservative rounded signal scales for translating environmental gradients into relative-uncertainty language.
It is noted that they are not claimed to reproduce HUST-99, UWash-00, or the BIPM balances, each of which would require its own field model and estimator-specific transfer function.

\subsection{Implications for future relative targets below $10^{-6}$ and long-correlation processes}
For a target relative level $u_{r,\rm target}$, Eq.~\eqref{eq:ur_env_grad} implies a simple requirement on the (estimator-standard) environmental gradient level,
\begin{equation}
\sigma_\Gamma \lesssim u_{r,\rm target}\,\Gamma_{\rm sig}.
\label{eq:sigma_requirement}
\end{equation}
This is a requirement (or design bound), not a new definition: to keep the environmental contribution below the chosen target relative uncertainty, the estimator-standard environmental gradient should not exceed the right-hand side.
The symbol $\lesssim$ is used because the benchmark signal gradients are rounded scales rather than apparatus-specific exact numbers.
For the benchmark signal gradients in Eq.~\eqref{eq:Gamma_sig_bench}, a target relative uncertainty of $10^{-6}$ corresponds to $\sigma_\Gamma\lesssim10^{-6}\Gamma_{\rm sig}$, i.e.\ roughly $10^{-13}$--$10^{-12}$~\si{\per\square\second}.
A target of $10^{-7}$ corresponds to $\sigma_\Gamma\lesssim10^{-7}\Gamma_{\rm sig}$, i.e.\ roughly $10^{-14}$--$10^{-13}$~\si{\per\square\second} over the same benchmark range.
These levels begin to overlap the high-noise/slow-advective atmospheric envelope and are plausibly reached (or exceeded) by local mass redistribution, hydrology, or building-specific pressure patterns.

It is also important to distinguish ``exceeding thermal noise'' from ``limiting $G$.''
Equation~\eqref{eq:sigmaGammaEq} addresses when environmental gradients exceed a thermal mean-torque reference for a chosen benchmark torsion balance and integration time.
Because $\sigma_{\Gamma,{\rm eq}}$ can be extremely small for high-sensitivity balances, it is possible for environmental fluctuations to dominate the stochastic torque uncertainty even when $u_{r,{\rm env}}(G)$ is far below a given target.

Finally, two cases should be distinguished clearly.
For stationary environmental inputs with long but finite correlation times, averaging saturates according to the OU benchmark of Sec.~\ref{sec:OUbenchmark} and the resulting contribution can still be propagated through Eq.~\eqref{eq:var_general}.
By contrast, true baseline wandering with no mean-reverting timescale on the run duration falls outside that benchmark and should be treated separately, as discussed in Sec.~\ref{sec:nonstationary}, typically through a dedicated drift model or as a Type-B component (or conservative Type-A bound) within the GUM framework.

\subsection{Comparison with the current CODATA relative standard uncertainty of $2.2\times10^{-5}$}
The current recommended value of $G$ carries a relative standard uncertainty of about $2.2\times10^{-5}$~\cite{CODATA2022,NISTCODATA}.
This number is not the uncertainty of any single apparatus; rather, it provides a useful reference scale for judging whether an environmental contribution is potentially relevant.

Setting $u_{r,\rm target}=2.2\times10^{-5}$ in Eq.~\eqref{eq:sigma_requirement} gives, for the benchmark signal gradients in Eq.~\eqref{eq:Gamma_sig_bench},
\begin{equation}
\sigma_{\Gamma,\mathrm{ref}} \approx 2.2\times10^{-5}\,\Gamma_{\rm sig}
\simeq
\begin{cases}
2\times10^{-12}\ \si{\per\square\second} & (\Gamma_{\rm sig}=10^{-7}\ \si{\per\square\second}),\\
2\times10^{-11}\ \si{\per\square\second} & (\Gamma_{\rm sig}=10^{-6}\ \si{\per\square\second}).
\end{cases}
\label{eq:sigma_22ppm}
\end{equation}
To relate Fig.~\ref{fig:pressure_to_torque} to an estimator-standard gradient level, we interpret each torque ASD curve as a one-sided spectrum in Hz and apply the simple-mean transfer function over an integration time $T_{\Sigma}$,
\begin{equation}
 u^2_{\rm atm}(\hat{\tau};T_{\Sigma})
 = \int_0^{\infty}{\rm d}f\,\mathrm{sinc}^2(\pi f T_{\Sigma})\,\mathrm{ASD}_{\tau}^2(f),
 \qquad
 \sigma_{\Gamma,\rm atm}(T_{\Sigma})=\frac{u_{\rm atm}(\hat{\tau};T_{\Sigma})}{|C_\Gamma|}.
\label{eq:sigmaGamma_fromASD}
\end{equation}
For the representative dumbbell used in Fig.~\ref{fig:pressure_to_torque}, $C_\Gamma=2ml^2\approx 2.65\times10^{-3}$~\si{kg.m^2}.
Evaluating Eq.~\eqref{eq:sigmaGamma_fromASD} for $T_{\Sigma}=10^{3}$--$10^{5}$~s across the four benchmark curves (NLNM/NHNM and acoustic/advective mappings) gives a full bracket from about $4\times10^{-18}$ to $1\times10^{-13}$~\si{\per\square\second}, with a representative central envelope around $10^{-16}$--$10^{-14}$~\si{\per\square\second}.
Within this benchmark comparison, standard atmospheric ``background'' pressure fluctuations are therefore not expected to be a limiting contributor at the current $2.2\times10^{-5}$ scale.

This conclusion should not be over-interpreted: local mass redistribution (nearby moving masses, hydrology, ventilation-driven pressure patterns, or building-specific flows) can exceed global-background benchmarks, and a claim about any specific apparatus requires its own transfer function and estimator.
Equation~\eqref{eq:sigma_requirement} provides a transparent way to translate a site-specific characterization of $\sigma_\Gamma$ into a bound on $u_{r,{\rm env}}(G)$.

\section{Mitigation and experimental design implications}
The expressions above suggest practical mitigation strategies:
\begin{itemize}
\item \emph{Symmetry with controlled anisotropy:} idealized quadrupole cancellation reduces long-wavelength coupling, but a practical $G$ pendulum must retain a measurable signal channel through controlled residual multipole content or matched higher-order source modulation.
\item \emph{Modulation:} shifting the measurement band away from the longest correlation times and separating stationary NN from slow drift.
\item \emph{Environmental sensing:} arrays of barometers/hydrology sensors can estimate local gradients and enable regression subtraction, as demonstrated in related NN subtraction studies~\cite{Coughlin2018}.
\item \emph{Drift management:} non-stationary baseline wandering should be diagnosed and bounded separately from the stationary environmental input, because its uncertainty treatment is not the same as that of the OU benchmark.
\end{itemize}

Within the estimator-based expression Eq.~\eqref{eq:var_general}, regression subtraction corresponds to choosing a weight function $w(t)$ that (after demodulation/averaging) subtracts a linear combination of auxiliary sensor channels.
For stationary noise this can be formulated in the frequency domain as a multichannel Wiener filter; the achievable suppression depends on sensor noise and on how well the array samples the relevant spatial scales set by $k^{-1}$.

\begin{figure}[t]
\centering
\includegraphics[width=0.85\linewidth]{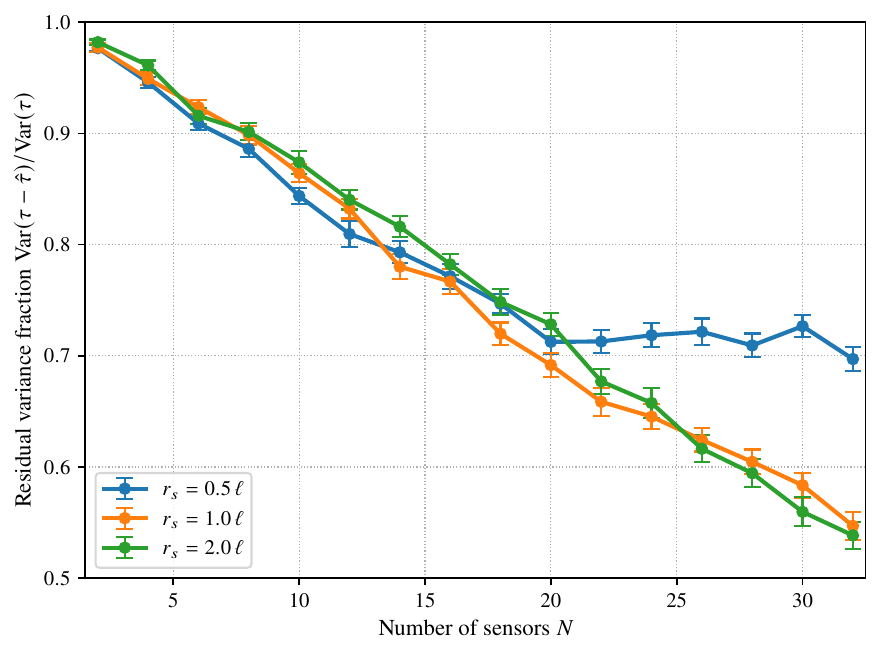}
\caption{Simplified Monte Carlo illustration of regression subtraction with a ring array of barometers.
Markers show the Monte Carlo mean over $N_{\rm MC}=50$ independent realizations; error bars indicate $\pm1$ standard error.
Shown is the residual variance fraction $\mathrm{Var}(\tau-\hat{\tau})/\mathrm{Var}(\tau)$ after multichannel Wiener subtraction, for $N$ sensors equally spaced on a ring of radius $r_s$.
The random field is generated as a superposition of plane-wave modes with a broad (log-uniform) $k$ distribution and additive sensor noise at the $\sim5\%$ level.
In this broad-band dimensionless toy model, arrays with $r_s$ of order $l$ to a few $l$ perform substantially better than the compact $r_s=0.5l$ array; for larger $N$, the $r_s=2l$ curve is comparable to or slightly below the $r_s=l$ curve.
The figure is qualitative only: because it omits atmospheric altitude weighting $e^{-k z_0}$ and any site-specific $P_\Sigma(k,\omega)$, it should not be interpreted as a quantitative prescription that the optimal real barometer radius is exactly $r_s=l$. Details are seen in Appendix~C.}
\label{fig:regression}
\end{figure}

To provide a compact numerical illustration of how regression can be incorporated into the estimator-based framework, we simulate a deliberately simplified isotropic surface-density field as a superposition of plane-wave modes with random directions, phases, and log-uniform wavenumbers $k$.
For each random-field realization, we compute the population covariances between the (dimensionless) dumbbell torque and $N$ idealized barometers uniformly distributed on a ring of radius $r_s$, and we evaluate the optimal linear estimator in the infinite-sample limit (equivalently, a multichannel Wiener subtraction).
Figure~\ref{fig:regression} shows the residual variance fraction as a function of $N$ for three sensor radii.
The purpose of this simulation is only to illustrate the regression algebra and the role of array geometry in a stationary field.
All lengths are normalized by the dumbbell half-arm length $l$, the mode band is sampled broadly over $kl\in[0.05,20]$, and no atmospheric altitude weighting $e^{-k z_0}$ or site-specific $P_\Sigma(k,\omega)$ is imposed.
Accordingly, Fig.~\ref{fig:regression} is not a quantitative prescription that real barometers should be placed at $r_s=l$.
Because the toy model is written in dimensionless form, the relevant control parameter is the ratio $r_s/l$ together with the weighted spectrum in $kl$; absolute metre-scale conclusions require a site-specific $P_\Sigma(k,\omega)$ and the kernel $e^{-k z_0}$.
Its robust message is more limited: a very compact array ($r_s\ll$ the dominant coupled spatial scale) is largely common-mode and subtracts poorly, whereas an array with radius of order the coupled spatial scale performs better.
In the present broad-band toy model, both $r_s=l$ and $r_s=2l$ outperform $r_s=0.5l$, and $r_s=2l$ becomes comparable to or slightly better than $r_s=l$ for the larger-$N$ points.
This illustration addresses only subtraction of a stationary environmental field; it is not intended as a model of non-stationary instrument drift.
Details of the toy model and all simulation parameters are given in Appendix~C and in the accompanying figure-generation code.

\section{Conclusion}
We developed an analytical framework for incorporating environmental gravity-gradient (Newtonian) noise into the uncertainty budget of torsion-balance measurements of $G$ within the measurement-model approach of the GUM.
The key ingredients are (i) estimator-dependent propagation of correlated torque noise via Eq.~\eqref{eq:var_general}, and (ii) closed-form transfer functions and baseline suppression factors that quantify the coupling from spatio-temporal surface-density fields to torque.
For the spatial coupling, the dumbbell and symmetric-cross geometries should be interpreted as bounding cases: the dumbbell is a transparent upper-coupling reference, whereas the perfectly symmetric cross is an idealized rejection benchmark.
Because exact quadrupole cancellation would also suppress any $G$ signal carried by the same low-order channel, a realizable experiment requires controlled residual anisotropy or a matched higher-order source modulation.
For temporal averaging, the OU model is used only as a stationary benchmark for the environmental input; it does not solve the separate problem of non-stationary baseline wandering in Cavendish-type measurements, which must be modeled or bounded independently.
Using the pressure NLNM/NHNM envelopes as a realistic atmospheric benchmark, we illustrated how spatial suppression and the assumed $k$--$\omega$ relation control the resulting torque spectra.
For representative laboratory heights ($z_0\sim\SI{1}{m}$), the simple-mean propagation of Fig.~\ref{fig:pressure_to_torque} corresponds to sample-mean equivalent gradients with a representative central envelope around $10^{-16}$--$10^{-14}$~\si{\per\square\second} for $T_{\Sigma}\sim10^{3}$--$10^{5}$~s (with the full benchmark bracket spanning more widely depending on the pressure level and the $k$--$\omega$ mapping); for the illustrative benchmark signal gradients in Eq.~\eqref{eq:Gamma_sig_bench} this remains well below the current CODATA-scale relative uncertainty.
Our main metrological message is therefore not that atmospheric background NN already dominates any specific present $G$ measurement.
Rather, within the benchmark signal-gradient range treated here it lies below the current CODATA-scale reference, while experiments are approaching a regime where it should be explicitly bounded, monitored, and potentially subtracted as systematic floors continue to decrease toward and below the $10^{-6}$ level.

\section*{Acknowledgement}
This work was partially supported by JST ASPIRE (No.~JPMJAP2339) and the JST SPRING Program (No.~JPMJSP2124).

\appendix
\section{Dumbbell baseline factor and small-$\alpha$ expansion}
Starting from the angular average,
\begin{equation}
A(\alpha)=\frac{1}{2\pi}\int_{0}^{2\pi}\sin^2\varphi\,\sin^2(\alpha\cos\varphi)\,{\rm d}\varphi,\qquad \alpha\equiv kl,
\end{equation}
use $\sin^2 x=\tfrac12[1-\cos(2x)]$ and $\sin^2\varphi=\tfrac12[1-\cos(2\varphi)]$ to write
\begin{equation}
A(\alpha)=\frac14-\frac{1}{4\pi}\int_{0}^{2\pi}\sin^2\varphi\,\cos(2\alpha\cos\varphi)\,{\rm d}\varphi.
\end{equation}
The remaining integrals can be evaluated with standard Bessel identities,
\begin{equation}
\int_0^{2\pi}\cos(2\alpha\cos\varphi)\,{\rm d}\varphi = 2\pi J_0(2\alpha),\qquad
\int_0^{2\pi}\cos(2\varphi)\cos(2\alpha\cos\varphi)\,{\rm d}\varphi = -2\pi J_2(2\alpha),
\end{equation}
yielding the closed form
\begin{equation}
A(\alpha)=\frac14\left[1-J_0(2\alpha)-J_2(2\alpha)\right].
\end{equation}
For $\alpha\ll1$, using
$J_0(2\alpha)\approx 1-\alpha^2+\alpha^4/4+\cdots$ and
$J_2(2\alpha)\approx \alpha^2/2-\alpha^4/6+\cdots$ gives
\begin{equation}
A(\alpha)\approx \frac{\alpha^2}{8}-\frac{\alpha^4}{48}+\cdots,
\end{equation}
so the leading coefficient is $1/8$.

\section{Cross baseline factor and leading-order scaling}
The integral definition~\eqref{eq:A_cross_def} admits a rapidly convergent series.
With $\alpha=kl$ and $f(\varphi)=\cos\varphi\,\sin(\alpha\sin\varphi)-\sin\varphi\,\sin(\alpha\cos\varphi)$, one can make the Fourier structure explicit using Jacobi--Anger expansions.
Starting from
\begin{align}
\sin(\alpha\sin\varphi) &= 2\sum_{m=0}^{\infty} J_{2m+1}(\alpha)\,\sin[(2m+1)\varphi],\\
\sin(\alpha\cos\varphi) &= 2\sum_{m=0}^{\infty}(-1)^m J_{2m+1}(\alpha)\,\cos[(2m+1)\varphi],
\end{align}
(using standard Fourier--Bessel expansions), and applying product-to-sum identities
\begin{align}
\cos\varphi\,\sin[(2m+1)\varphi] &= \tfrac12\left[\sin((2m+2)\varphi)+\sin(2m\varphi)\right],\\
\sin\varphi\,\cos[(2m+1)\varphi] &= \tfrac12\left[\sin((2m+2)\varphi)-\sin(2m\varphi)\right],
\end{align}
one finds that only harmonics consistent with the fourfold symmetry survive.
A short reindexing of the even/odd $m$ terms yields
\begin{equation}
 f(\varphi)=2\sum_{n=1}^{\infty}\left[J_{4n-1}(\alpha)+J_{4n+1}(\alpha)\right]\sin(4n\varphi),
\end{equation}
and therefore
\begin{equation}
A_\times(\alpha)=\frac{1}{2\pi}\int_0^{2\pi} f(\varphi)^2\,{\rm d}\varphi
=2\sum_{n=1}^{\infty}\left[J_{4n-1}(\alpha)+J_{4n+1}(\alpha)\right]^2.
\end{equation}
For $\alpha\ll1$ the leading term is $n=1$, dominated by $J_3(\alpha)\approx \alpha^3/48$ (while $J_5(\alpha)=\mathcal{O}(\alpha^5)$), giving
\begin{equation}
A_\times(\alpha)\approx 2\,J_3(\alpha)^2\approx 2\left(\frac{\alpha^3}{48}\right)^2
=\frac{\alpha^6}{1152}+\cdots,
\end{equation}
so the leading coefficient is $1/1152$ and the long-wavelength scaling is $\propto (kl)^6$.

\section{Regression illustration of Figure~\ref{fig:regression}}
This appendix summarizes the simplified regression simulation used for Fig.~\ref{fig:regression}.
Its purpose is to provide a transparent, reproducible illustration of how auxiliary environmental channels can be incorporated into the estimator-based uncertainty propagation via linear regression (multichannel Wiener subtraction).

\paragraph{Random-field model.}
We represent an isotropic surface field as a superposition of many plane-wave modes with random directions $\theta_j$, phases $\phi_j$, and wavenumbers $k_j$ sampled log-uniformly over a broad band,
\begin{equation}
 x(\mathbf{r})=\sum_{j=1}^{N_{\rm modes}} a_j\cos\!\left(k_j\,\hat{\mathbf{k}}_j\cdot\mathbf{r}+\phi_j\right),\qquad
 \hat{\mathbf{k}}_j=(\cos\theta_j,\sin\theta_j),
\end{equation}
with independent standard-normal amplitudes $a_j$.
All quantities are made dimensionless by measuring lengths in units of the torsion-balance arm length $l$.

\paragraph{Toy torque coupling.}
For each mode, we use a deliberately simplified long-wavelength dumbbell-like coupling
\begin{equation}
 y\equiv \tau \propto \sum_{j=1}^{N_{\rm modes}} a_j\,g_j,\qquad g_j = \sin\theta_j\,\sin(k_j\cos\theta_j),
\end{equation}
chosen to mimic the angular dependence of Eq.~\eqref{eq:tau_dumbbell_k} while keeping the simulation algebraically transparent.

\paragraph{Sensor array and regression.}
We place $N$ idealized sensors uniformly on a ring of radius $r_s$ (in units of $l$), at positions
$\mathbf{r}_i=r_s\,(\cos(2\pi i/N),\sin(2\pi i/N))$.
The sensor readouts are
\begin{equation}
 x_i = x(\mathbf{r}_i)+n_i,
\end{equation}
where $n_i$ is independent sensor noise.
We implement a simple relative-noise model by setting $\mathrm{Var}(n_i)=\epsilon^2\,\mathrm{Var}[x(\mathbf{r}_i)]$ with $\epsilon=0.05$.
For each random-field realization, we compute the population (infinite-sample) covariances $\Sigma_x=\langle\mathbf{x}\mathbf{x}^T\rangle$ and $\mathbf{C}=\langle\mathbf{x}\,y\rangle$, and evaluate the optimal linear estimator $\hat{y}=\mathbf{b}^T\mathbf{x}$ with $\mathbf{b}=\Sigma_x^{-1}\mathbf{C}$.
The resulting residual-variance fraction is
\begin{equation}
\frac{\mathrm{Var}(y-\hat{y})}{\mathrm{Var}(y)}=1-\frac{\mathbf{C}^T\Sigma_x^{-1}\mathbf{C}}{\mathrm{Var}(y)}.
\end{equation}
The plotted error bars are Monte Carlo standard errors, $\mathrm{SEM}=s/\sqrt{N_{\rm MC}}$, where $s$ is the sample standard deviation of the residual fraction over the independent realizations.
They quantify finite-ensemble simulation uncertainty, not a physical instrumental error budget.

\paragraph{Parameters and scope.}
Figure~\ref{fig:regression} uses $N_{\rm MC}=50$ independent realizations with $N_{\rm modes}=120$ modes each, $k\in[0.05,20]$ sampled log-uniformly in units of $1/l$, and $\epsilon=0.05$.
The sensor radius is varied over $r_s/l\in\{0.5,1.0,2.0\}$ and the sensor count over $N=2,4,\ldots,32$.
The log-uniform sampling is intentional: it gives comparable weight to each decade in $k$ and therefore avoids hard-wiring a preferred spatial scale into this dimensionless toy model.
The chosen band $kl\in[0.05,20]$ spans wavelengths from $\lambda\approx125\,l$ down to $\lambda\approx0.3\,l$, so the simulation covers both the common-mode regime ($kl\ll1$) and the spatially resolved regime around and above $kl\sim1$.
For exact reproducibility, the supplied script fixes the random seed and can export the numerical means and standard errors used in the plot.
This appendix should be read together with one important limitation: the simulation does \emph{not} include the atmospheric gravity kernel $e^{-k z_0}$, the pressure NLNM/NHNM spectra, or any site-specific spatio-temporal model.
It therefore tests the formal regression framework rather than a literal atmospheric sensor-placement problem.
Because all lengths are normalized by $l$, Fig.~\ref{fig:regression} depends only on the dimensionless ratios $r_s/l$ and $kl$.
Its message is therefore not that there exists a universal absolute sensor radius independent of the apparatus; rather, the preferred array size tracks the dominant \emph{coupled spatial scale} after the real weighting by $e^{-k z_0}$ and the actual $P_\Sigma(k,\omega)$ are taken into account.
If one imposed a spectrum dominated by wavelengths much larger than $l$, all three curves would move upward (poorer subtraction) and their dependence on $r_s/l$ would become weaker; the preferred radius would then be set by the dominant weighted wavelength, not by $l$ alone.
The full implementation is provided in the accompanying figure-generation scripts.

\end{document}